\documentclass{aa}  

\usepackage{graphicx}
\usepackage[]{hyperref}
\usepackage{txfonts}
\usepackage{amsmath}
\usepackage{amsfonts}
\usepackage{makecell}
\usepackage{subcaption}         % necessary for continued figures, example in section 3
\usepackage{lscape}             % to rotate a single page table, example in appendix.
\usepackage{placeins}           % useful with \FloatBarrier, to keep 
\usepackage{booktabs}
\usepackage{natbib}
\usepackage[toc,page]{appendix}
\usepackage{xcolor}
                 
\hypersetup{
    colorlinks=true,
    citecolor=blue,
    linkcolor=blue,
    filecolor=magenta,      
    urlcolor=cyan,
}

\begin{document}

%%%%%%%%%%%%%%%%%%%%%%%%%%%%%%%%%%%%%%%%
% if you use custom commands in your title,
% ensure to check your title when submitting!
%%%%%%%%%%%%%%%%%%%%%%%%%%%%%%%%%%%%%%%%
   \title{Testing Refracted Gravity with the kinematics of stacked galaxy clusters}

   %\subtitle{Subtitle}

%%%%%%%%%%%%%%%%%%%%%%%%%%%%%%%%%%%%%%%%
% Please separate each author with the \and command
%
% Please do not include ORCIDs next to author names.
% Only ORCIDs authenticated by individual authors in EDPS
% editorial system will be taken into account.
% ORCIDs included here will be removed.
%%%%%%%%%%%%%%%%%%%%%%%%%%%%%%%%%%%%%%%%

   \author{D.Primignani \inst{1}\and L.Pizzuti \inst{1} \and A. Diaferio \inst{2,3,4} \and
   V. Broccolato\inst{2}}

   \institute{ Dipartimento di Fisica G. Occhialini, Università degli Studi di Milano-Bicocca, Piazza della Scienza 3, I-20126 Milano, Italy \and  Dipartimento di Fisica, Università di Torino, via P. Giuria 1, 10125, Torino, Italy \and Istituto Nazionale di Fisica Nucleare, sezione di Torino, via P. Giuria 1, Torino, 10125, Italy. \and Accademia delle Scienze di Torino, via Maria Vittoria 3, Torino, 10123, Italy.}

   \date{}

% \abstract{}{}{}{}{}
% 5 {} token are mandatory
 
\abstract{
We test Refracted Gravity (RG), a phenomenological modified gravity model that modifies the Poisson equation by introducing an effective gravitational permittivity $\epsilon(\rho_{\rm bar})$ which depends on the baryonic matter density $\rho_{\rm bar}$ and mimics the behaviour of dark matter in cosmic structures. We perform a kinematical analysis of member galaxies for a sample of 74 galaxy clusters, compiled from the Cluster Infall Regions in the Sloan Digital Sky Survey (CIRS). We stack the clusters in three bins of increasing velocity dispersion to mitigate the effects of systematic errors induced by deviations from dynamical equilibrium and from the assumption of spherical symmetry.
By means of the \texttt{MG-MAMPOSSt} code, we reconstruct
the gravitational potential in RG and infer
the model parameters jointly with the gas and stellar mass-profile parameters, adopting broad, physically informed
priors on the baryonic components. The estimates
of the main parameter in Refracted Gravity, the vacuum permittivity $\epsilon_0$, range from
$0.04$ to $0.37$, across the three main stacks,
depending on the model adopted for the hot gas component. The constraints are mutually compatible within the quoted uncertainties and overlap with previous cluster-scale estimates. The agreement becomes less sensitive to gas parametrisation when selecting the less-disturbed clusters for building the stacks.
These results characterize the sensitivity of cluster-scale
RG constraints to baryonic modelling and sample selection.}

\maketitle

\section{Introduction}

The standard cosmological model, or $\Lambda$CDM model, is based on General Relativity (GR) as the theory of gravity, and it currently represents the best description of our universe, efficiently matching several independent observations over a wide range of scales \citep[][]{2013AAS...22110601C,Heavens_2017,refId0}. Nonetheless, its success relies on introducing two non-baryonic dark components whose physical nature remains unknown. The first is cold dark matter (CDM), a form of collisionless, non luminous matter which constitutes up to 25\% of the total cosmic energy budget and is required to describe the formation and growth of structures \citep[e.g.,][]{dodelson1996cold,Freese_2009}. 
The second component, dark energy (DE), is required to explain the late-time accelerating expansion of the universe \citep{1998AJ....116.1009R,Perlmutter_1999}. In the context of the $\Lambda$CDM paradigm, DE takes the form of a cosmological constant $\Lambda$ in the Einstein's field equations of GR \citep[see e.g][]{Weinberg89, 2015daen.book.....A}. DE can be considered as an effective fluid with an equation of state (EoS) $p = w\rho c^2$, where $w = -1$, accounting for $\sim 70\%$ of the cosmic energy budget \citep[e.g.,][]{refId0}.

In the last decade, the increasing precision and accuracy of observations have posed additional challenges to the $\Lambda$CDM paradigm, identifying several inconsistencies between the prediction of the model and the observations, as well as tensions between local and early-time datasets \citep[e.g.,][]{Perivolaropoulos_2022,Abbott_2022,2025jwst.prop.7184M}. 
Moreover, recent observations conducted with the Dark Energy Spectroscopic Instrument (DESI, \citealt{Abdul_Karim_2025,Adame_2025}) revealed a statistically significant preference for a dynamically evolving DE with respect to the cosmological constant, although some studies argued that this evidence may be the consequence of systematic errors in the combination of datasets \citep{ong2026}.
 
Furthermore, several studies on galaxies found  empirical relations, such as the baryonic Tully-Fisher relation \citep{1977A&A....54..661T,2000ApJ...533L..99M}, the mass discrepancy and the radial acceleration relation \citep{2004ApJ...609..652M,PhysRevLett.117.201101}, that are not easily accounted for by the $\Lambda$CDM model. These relations exhibit a common acceleration scale $a_0\simeq 1.2 \times 10^{-10} \text{m}/\text{s}^{2}$ above which the presence of DM appears to be unnecessary. Intriguingly, $a_0 \sim c^2\sqrt{\Lambda}$, suggesting a possible link between DM and DE.

An interesting possibility to solve these issues is to modify GR on large length scales. This way, part of (or all)  the phenomenology attributed to the dark components can instead arise from modified gravity (MG), reproducing astrophysical and cosmological observations (see e.g., \citealt{2017PhR...692....1N} for a review).  In recent years many modified gravity models have been developed as alternative to DM \citep[e.g.][]{Clifton_2012,Shankaranarayanan_2022}. For example, Modified Newtonian Dynamics (MOND, \citealt{1983ApJ...270..365M}) encompasses a family of models where the theory of gravity is modified at accelerations below the characteristic scale $a_0$. While MOND successfully reproduces several observed phenomena and scaling relations at galactic scales, an additional unseen mass component is generally still required at larger scales \citep[][]{2012LRR....15...10F,2019MNRAS.485..513S,Kelleher:2024oip}.

 In this work, we focus on Refracted Gravity (RG hereafter, \citealt{2016arXiv160304943M}), a phenomenological MG model where the Poisson equation is modified by introducing a gravitational permittivity which depends on the (baryonic) matter density. %, in analogy with electrodynamics.  
 This new term enhances the gravitational field when the density is below a critical value, resulting in a modification of the gravitational potential that could mimic the effect of DM at both galaxy and galaxy cluster scales \citep[e.g.,][]{2024Astro...3...68C}.
A covariant version of RG has been recently proposed \citep{2023A&A...674A.209S}:  RG is treated as a scalar-tensor theory, where a single scalar field is added to the action to introduce the gravitational permittivity. In this formulation, it turns out that the same scalar field mimicking DM is also responsible for a new term at low densities that could describe the accelerating expansion of the universe. However, recent analyses \citep{2026JCAP...06..038G} indicate that this specific covariant RG model is not able to form cosmic structure at sufficiently early times as required by observations. Nevertheless, it remains an interesting test-bed of theories providing a unified description of the dark sector. 

Several tests of the phenomenological RG have been performed at different scales: from disk and elliptical galaxies \citep{2020A&A...637A..70C,2022A&A...657A.133C} to galaxy clusters \citep{2025A&A...698A..83P}. Among different probes employed to test the gravitational interaction at large scales, galaxy clusters occupy a privileged role. They are the most massive cosmic structures that have undergone gravitational collapse, probing the peaks of the initial density perturbations and growing via hierarchical mergers and accretion. According to $\Lambda$CDM, the majority of the matter budget ($\sim 80\%$) in  clusters is made by CDM, which dominates the overall gravitational dynamics, while the internal properties are strongly influenced by baryonic physics. This interplay makes galaxy clusters an excellent laboratory for studying processes at the edge between astrophysics and cosmology.

\cite{2025A&A...698A..83P} performed a detailed kinematical analysis in RG of two massive clusters within the CLASH-VLT \citep{Rosati2014} sample, using high quality data of line-of-sight velocities and projected positions of member galaxies, combined with precise measurements of the stellar line-of-sight velocity dispersion of the brightest cluster galaxy (BCG). These joint datasets provided tight bounds on the free parameters of RG; however, the constraints are not in agreement between the two clusters, and exhibit a further tension with previous results obtained on galaxy scale (e.g., \citealt{2022A&A...657A.133C}). %%%%%%%% 
On the one hand, these results can be interpreted as a signal of the failure of the phenomenological RG model; on the other hand, the origin of this tension could be attributed to systematic effects in the analysis, rather than to fundamental issues with the theory.  In particular, the kinematical modelling of \cite{2025A&A...698A..83P} is based on the assumption that galaxy clusters are spherically symmetric and in dynamical equilibrium. In fact, the modification of the Poisson equation in RG is strongly dependent on the shape of the matter distribution \citep{2016arXiv160304943M}, and deviation from a regular morphology or equilibrium configuration may significantly bias the constraints, especially when the statistical uncertainties are very small. 

In this work, we aim to mitigate the effects of such systematic errors and to understand the nature of the tension by considering a large sample of clusters (74 clusters) from the Cluster Infall Regions in the Sloan Digital Sky Survey (CIRS,  \citealt{2006AJ....132.1275R}). In order to reduce the impact of deviation from a spherical symmetry, we perform a stacking in 3 bins of the galaxy velocity dispersion $\sigma$, which is a proxy of the total dynamical mass of the cluster \citep[see e.g.][]{2008ApJ...672..122E}. We employ the \texttt{MG-MAMPOSSt} code of \cite{2021MNRAS.506..595P} to model the gravitational potential in RG and the velocity anisotropy profile of the stacked clusters in all the bins. We constrain the gravitational permittivity in vacuum $\epsilon_0$ and compare the results across different bins and with previous work to assess the consistency of the inferred vacuum permittivity and its sensitivity to baryonic
modelling and sample selection.

This paper is structured as follows: in Sect.~\ref{sec:RG}, we review the theoretical framework and the main equation of RG; in Sect.~\ref{sec:data}, we describe our dataset. In Sect.~\ref{sec:mgmamposst}, we provide a brief overview of the code \texttt{MG-MAMPOSSt} and  how it is used to perform the analysis, while in Sect.~\ref{sec:results} we present our results, which are further discussed in Sect. \ref{sec:discussion}. Finally, in Sect.~\ref{sec:conclusion}, we draw our conclusions.  In the following, where required, we adopt the standard cosmological background with $\Omega_{\Lambda}=0.7$, $\Omega_m=0.3$ and $H_0=70  \,\text{km} \,\text{s}^{-1}\text{Mpc}^{-1}$.

\section{Refracted Gravity}
\label{sec:RG}
Refracted gravity is a phenomenological MG model developed by \cite{2016arXiv160304943M} that introduces a gravitational permittivity $\epsilon(\rho_{\rm bar})$ as a monotonic function of the baryonic mass density $\rho_{\rm bar}$. In this model, the gravitational potential $\Phi$ behaves in analogy to the electromagnetic fields when passing from a medium with permittivity $\epsilon_1$ to another with permittivity $\epsilon_2$: the field lines are refracted and both magnitude and direction of the field could change. This modification allows for a magnification of the gravitational potential at large scales without requiring a DM component. 
The introduction of the gravitational permittivity modifies the Poisson equation as
\begin{equation}
    \nabla\cdot\left[\epsilon(\rho_{\rm bar})\nabla \Phi\right]=4\pi G\rho_{\rm bar} \,,
    \label{RG Poisson equation}
\end{equation}
where $G$ is the Newtonian gravitational constant. Eq.~\eqref{RG Poisson equation} can be rewritten as 
\begin{equation}
   \frac{\partial \epsilon}{\partial \rho_{\rm bar}}\nabla\rho_{\rm bar}\cdot\nabla \Phi+\epsilon\nabla^2\Phi=4\pi G\rho_{\rm bar}\,.
   \label{RG Poisson equation 2}
\end{equation}
The above expression highlights two related effects introduced by the gravitational permittivity. The first term, which arises from spatial variations of $\epsilon$, couples the gravitational field to gradients of the baryonic density and can modify both the magnitude and the geometry of the field; in non-spherical systems, this produces the characteristic refraction of the field lines. The factor $\epsilon$ multiplying the standard Poisson term instead modifies the strength of the gravitational response: where $\epsilon<1$, a given baryonic density generates a stronger gravitational field than in Newtonian gravity.
The permittivity $\epsilon$ is assumed to be 1 for large densities, recovering the original  Poisson equation at small scales, where GR is well tested, and equal to $\epsilon_0 <1$ in vacuum and at large scales, where the density is much lower than a critical density $\rho_c$:
\begin{equation}
    \epsilon(\rho_{\rm bar})\simeq
    \begin{cases}
        1&\rho_{\rm bar}\gg\rho_c\\
        \epsilon_0&\rho_{\rm bar}\ll\rho_c
    \end{cases}\,.
    \label{approx_epsilon}
\end{equation}
The functional form of the permittivity is assumed to be a smooth step function of the mass density alone,
\begin{equation}
    \epsilon(\rho_{\rm bar})=\epsilon_0
    +(1-\epsilon_0)\frac{1}{2}\left\{\tanh\left[Q\,\ln\left(\frac{\rho_{\rm bar}}{\rho_c}\right)\right]+1\right\}
    \label{epsilon (rho)} \,,
\end{equation}
where $Q$ is a parameter that controls the steepness of the transition between the high- and low-density regimes. %\LP{Forse andrebbe detto che è una forma fenomenologica?}

If we consider spherically-symmetric systems,  the field lines remain radial, as $\nabla\rho_{\rm bar}$ and $\nabla \Phi$ are parallel. At high densities, where $\rho_{\rm bar}\gg\rho_c$ and $\epsilon\simeq1$, the standard Newtonian behaviour is recovered. Conversely, at low densities, where $\rho_{\rm bar}\ll\rho_c$ and $\epsilon\simeq\epsilon_0<1$, the gravitational acceleration generated by a given enclosed baryonic mass is enhanced by a factor $1/\epsilon_0$ relative to its Newtonian value. If sufficiently large, this enhancement may reproduce the gravitational effect conventionally attributed to CDM and thereby account for the observed velocities using the baryonic mass alone.

Under spherical symmetry, the modified Poisson equation becomes
\begin{equation}
\frac{1}{r^2}\frac{\mathrm{d}}{\mathrm{d}r}
\left\{
r^2\epsilon\bigl[\rho_{\rm bar}(r)\bigr]
\frac{\mathrm{d}\Phi}{\mathrm{d}r}
\right\}
=
4\pi G\rho_{\rm bar}(r).
\end{equation}
Integrating this equation over a sphere of radius $r$ gives
\begin{equation}
\frac{\mathrm{d}\Phi}{\mathrm{d}r}
=
\frac{G M_{\rm bar}(<r)}
{r^2\epsilon\bigl(\rho_{\rm bar}(r)\bigr)},
\label{Gradient Phi}
\end{equation}
where $M_{\rm bar}(<r)$ is the baryonic mass enclosed within $r$, and $\epsilon(\rho_{\rm bar})$ is given by Eq.~\eqref{epsilon (rho)}. In this case, the total baryonic density is mostly (80\%) composed  by the hot X-ray emitting gas of the intra-cluster medium (ICM) and by the stellar mass in member galaxies, $\rho_{\rm bar}(r)=\rho_{\rm gas}(r)+ \rho_{\rm gal}(r)$, plus a diffuse stellar component, the intra-cluster light (ICL), which typically contributes only a subdominant fraction of the total baryonic budget, of order a few per cent in massive clusters \citep[e.g.,][]{Gonzalez2013}. We will not consider the explicit ICL contribution here. The corresponding total baryonic mass is therefore obtained as 
 \begin{equation}
     M_{\rm bar}(r)=4\pi\int_0^r \text{d}x\, \rho_{\rm bar}(x)x^2 = M_{\rm gas}(r)+M_{g\rm al}(r) \,.
 \end{equation}
Close to the centre ($r \lesssim50 $ kpc) the mass is dominated by the BCG, typically a giant elliptical galaxy sitting at the minimum of the gravitational potential; analyses aiming at reconstructing the mass profile down to the cluster core have to separate the BCG component and explicitly model its profile (e.g. \citealt{2025A&A...698A..83P}). In the present work we do not include an independent BCG mass component and we limit our analysis to $r\gtrsim 50$ kpc.
This radial cut reduces the direct sensitivity of the kinematical data
to the internal BCG profile, but it does not formally remove the
contribution of its enclosed mass at larger radii. Neglecting this
component therefore remains an approximation of the present analysis,
as discussed in Sect.~\ref{sec:discussion}.

\section{Dataset}
\label{sec:data}
We use a sample of clusters from the Clusters Infall Regions in SDSS (CIRS, \citealt{2006AJ....132.1275R}) catalogue.  This sample was obtained by relying on the photometric and spectroscopic data from the fourth data release of the Sloan Digital Sky Survey (SDSS),  a large survey covering almost one fourth of the sky at high galactic latitudes.  This survey  principally contains low-redshift galaxies, with a limit of $z\leq0.10$, %implying that the clusters in the sample are near and massive, 
with cluster masses, estimated in Newtonian gravity, between $10^{13}$ and $10^{15}\, \text{M}_{\odot}$. X-ray fluxes are listed in X-ray catalogues such as REFLEX and NORAS \citep{2002ApJ...566...93B,2000ApJS..129..435B}, and the sample is flux-limited with a minimum flux criterion in the X-band of $L_X\geq3\times10^{-12}\text{erg s}^{-1}\text{cm}^{-2}$. 
 The main advantage of this sample is the large number of clusters and the wide range in masses and X-ray luminosities that it covers.

The CIRS sample contains 74 clusters and a total of 17798 galaxies with spectroscopic redshift, with 7601 galaxies within the cluster virial radii. 
\cite{2006AJ....132.1275R} used the caustic technique \citep{1997ApJ...481..633D,1999ApJ...517L..23G} to study the mass profile of these clusters and estimated the virial radius $r_{200}$ from the mean enclosed density profile. Here and below, we indicate with $r_{\Delta}$   the radius enclosing a spherical overdensity $\Delta$ times the critical density of the universe at that redshift $\rho_c(z)$; we denote the corresponding mass $M_\Delta$. 
\cite{2006AJ....132.1275R} further found that the mass profiles are well described by a Navarro-Frenk-White (NFW, \citealt{1997ApJ...490..493N}) and Hernquist \citep{1990ApJ...356..359H} models and that data disfavour the singular isothermal sphere model (SIS).

\subsection{Binning and stacking}
We divide the data sample into three bins in velocity dispersion, which is proportional to the total mass of the cluster (see e.g. \citealt{2008ApJ...672..122E, 2021ApJ...910...56T,2025A&A...699A..88P}). 
In our analysis, we use the spectroscopic information of the clusters' projected phase space (p.p.s hereafter), i.e the projected positions ($R$) and line-of-sight velocities of each member galaxy ($v_z$), computed in the cluster's rest frame.
We combine the p.p.s distributions of the clusters in physical coordinates, without rescaling the projected radii and line-of-sight velocities by the individual values of $r_{200}$ and $v_{200}$. The resulting stacks should consequently be regarded as effective ensemble systems retaining the main kinematical properties of their constituents.

 The first bin contains clusters with galaxy velocity dispersion $\sigma \lesssim 550$ km/s. The bin contains 35 clusters with a total of 5196 galaxies.
The second bin contains 25 clusters with $\ 550<\sigma<750$ km/s, which amounts to a total of 6354
galaxies, whereas the last bin contains 14 massive clusters with $\sigma \gtrsim 750$ km/s, for a total of 6248 galaxies.  The characteristic and physical properties of each bin, including the mean values of  $r_{200}$ and $M_{200}$ estimated in Newtonian gravity, are summarized in Table \ref{tab:Binning}.
After separating the clusters into the three bins described above, we performed the stacking of the phase space of all the clusters in each bin.

\begin{figure*}
\centering
\includegraphics[width=0.28\textwidth]{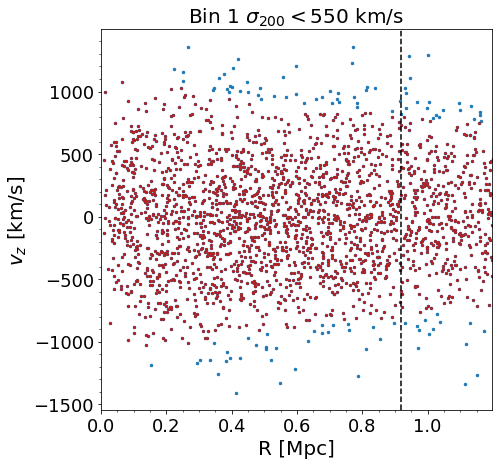}
\includegraphics[width=0.28\textwidth]{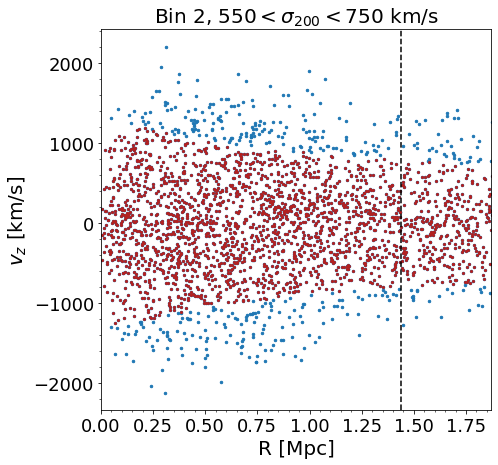}
\includegraphics[width=0.28\textwidth]{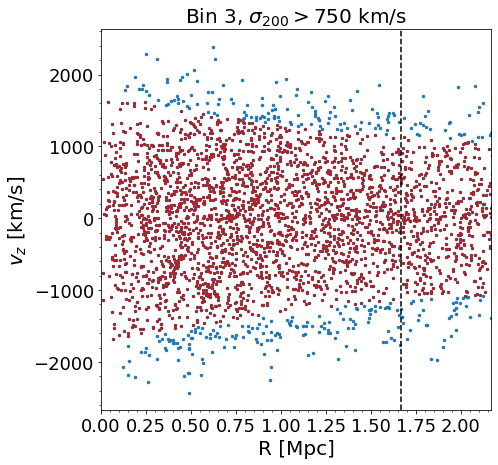}
\caption{Plot of the stacked projected phase spaces of clusters in the three bins. Galaxies selected by the Clean algorithm are coloured in red, while interlopers are coloured in blue. The dotted vertical line corresponds to the mean value of $r_{200}$ in each bin.}
\label{Figure p.p.s}
\end{figure*}

While stacking is an efficient way to reduce effects of triaxiality in clusters, by averaging over different orientations \citep{van_der_Marel_2000}, other systematic effects related to departures from dynamical equilibrium and to the choice of the cluster centre for each system in the stacked bin may still give rise to relevant biases \citep{Cai2025}. Moreover, residual interlopers -- i.e., galaxies within the projected cylinder $R\lesssim r_{200}$ which do not belong to the cluster (three-dimensional distance from the cluster centre $r_{\rm gal} \gg r_{200}$) -- can produce a significant contamination in the stacked cluster. As such, we run the \texttt{Clean} algorithm \citep{2013MNRAS.429.3079M} on each stacked p.p.s, a method which selects cluster members by an iterative process based on the estimate of the line-of-sight velocity dispersion to progressively remove interlopers. 
The stacked p.p.s are plotted in Fig.~\ref{Figure p.p.s}, where we have marked in red the final member selection after \texttt{Clean}. We retain 1909, 1768, and 2337 members within $r_{200}$ for Bin 1, Bin 2, and Bin 3, respectively. 

In order to further reduce the effect of un-relaxed systems, we employ a simple criterion to select a subsample of galaxy clusters showing indications of dynamical equilibrium, which is one of the key assumptions of the kinematical analysis implemented in the \texttt{MG-MAMPOSSt} code.
The distribution of velocities in a cluster constitutes an important indication of its dynamical state. As a matter of fact, if the velocity distribution along the line-of-sight is compatible with a Gaussian centred on 0,  galaxies are in random orbits around the centre of mass of the cluster with a symmetric velocity distribution, which is a signal of dynamical relaxation \citep{2013A&A...556A..74R}. 
In this regard, a good proxy of how well the velocity distribution is compatible with a Gaussian probability density is the Anderson-Darling coefficient $A^2$ \citep{anderson_darling_1952}, which has been shown to strongly correlate with the systematic bias induced by un-relaxed clusters \citep{Pizzuti2020syst}. In particular, large values of $A^2$ (corresponding to large departure from Gaussianity) result in a higher probability to get a biased kinematic mass reconstruction. 
We thus compute $A^2$ for all the individual clusters in our sample by considering galaxies within the corresponding $r_{200}$. 

We then re-compute new stacked p.p.s by selecting only galaxy clusters with a value of the Anderson-Darling coefficient lower than 0.7. In this case, to keep enough statistics per bin, we consider only two groups in velocity dispersion: ``Bin 1 relaxed" and ``Bin 2 relaxed". The first one considers clusters with velocity dispersion $\sigma \leq 600\,\text{km/s}$, and contains a total of 4774 galaxies. The second one includes clusters with $\sigma >600\, \text{km/s}$, for a total of 7207 galaxies. The main physical properties of these new bins are also reported in Table \ref{tab:Binning}.

\begin{table*}
\centering
{\small
\setlength{\tabcolsep}{4pt}
\renewcommand{\arraystretch}{0.95}
\begin{tabular}{lccccccccc}
\toprule
Bin &
\makecell{$\sigma$\\$[\mathrm{km\,s^{-1}}]$} &
$N_{\rm cl}$ &
$N_{\rm gal}$ & \makecell{$N_{\rm gal}$ \\after Clean} & 
$\bar{z}$ &
\makecell{$\bar{r}_{200}$\\$[\mathrm{Mpc}]$} &
\makecell{68th percentile range\\$[\mathrm{Mpc}]$} &
\makecell{$\bar{M}_{200}$\\$[10^{14}\,\mathrm{M}_{\odot}]$}
& \makecell{68th percentile range\\$[10^{14}\,\mathrm{M}_{\odot}]$}\\
\midrule
1         & $\sigma<550$       & 35 & 5196 & 4715&  0.048 & 0.92 & [0.66-1.15] & 1.1 & [0.35-1.86]\\
2         & $550<\sigma<750$   & 25 & 6354 & 5476 & 0.068 & 1.49 &[1.25-1.62] & 3.8 & [2.29-5.17] \\
3         & $\sigma>750$       & 14 & 6248 &5655 & 0.074 & 1.62 & [1.51-1.89] & 5.9 & [4.27-8.27] \\
1 relaxed & $\sigma<600$       & 33 & 4774 &  4161 & 0.050 & 0.95 & [0.67-1.21]  & 1.3 & [0.34-2.21] \\
2 relaxed & $\sigma>600$       & 22 & 7207 & 6232 & 0.065 & 1.57 & [1.38-1.78] & 4.9 & [3.3-6.92] \\
\bottomrule
\end{tabular}
}
\caption{Main properties of the velocity-dispersion bins. Barred quantities represent the mean values within each bin.}
\label{tab:Binning}
\end{table*}

\subsection{Gas and galaxies mass profiles}
\label{sec:gas}
 In RG the gravitational potential is sourced by baryonic matter alone (gas and galaxies) via Eq.~\eqref{RG Poisson equation 2}. The appropriate modelling of the RG potential would thus require an accurate modelling of the distribution of the baryonic matter. Unfortunately, we were able to retrieve data from the literature for the electron density profile to obtain the total ICM mass only for a few clusters in our sample, whereas the stellar mass profile is unknown for all the systems. Therefore, we adopt a simplified, quite general approach: by assuming parametric models for both $M_{\rm gas}$ and $M_{\rm gal}$, we let the kinematic data to fit the average gas and stellar profiles in the stacked clusters, along with the RG free parameters. As commonly done, we describe the gas density of the ICM with a beta profile \citep{1976A&A....49..137C},
\begin{equation}
    \rho_{\rm gas}(r)=\frac{\rho_{\rm g}}{\left[ 1+\left(\frac{r}{r_{\rm g}}\right)^2\right]^{\frac{3}{2}\beta}},
    \label{Beta model}
\end{equation}
where $\rho_{\rm g}$ is the gas density in the centre of the cluster, $r_{\rm g}$ is a scale radius of the model and $\beta$ indicates the characteristic slope. In our analysis we adopt two choices for the exponent $\beta$: the standard isothermal choice $\beta = 2/3$, and a more rapidly-decreasing density profile with $\beta = 1$. For the purpose of our paper, we do not consider more sophisticated models to avoid over fitting; we have indeed checked in preliminary analyses that the double-beta profiles are disfavoured by the Bayesian Information Criterion (BIC hereafter, \citealt{1978AnSta...6..461S}, see Sect.~\ref{sec:results}). 
The galaxy stellar mass density profile is parametrised according to a NFW model \citep{1997ApJ...490..493N}, whose density reads
\begin{equation}
    \rho_{\rm gal}(r) = \frac{\rho_\star}{\frac{r}{r_\star }\left(1+\frac{r}{r_\star }\right)^2}\,,
\end{equation}
with $\rho_{\star}$ the characteristic density and $r_{\star}$ the scale radius where the logarithmic slope of the density is equal to $-2$. This profile has been shown to provide a good fit for both the mass and number density profile of member galaxies in clusters \citep[e.g.,][]{10.1093/mnras/sty073,Biviano_2023}.
We anticipate here that $\rho_{\rm gal}(r)$
should not be confused with the number density profile $\nu(r)$ of the
galaxies used as kinematical tracers in the Jeans equation (see Sect.~\ref{sec:mgmamposst}). By denoting by
$\phi(M_\star,r)$ the radial stellar-mass function of the selected
galaxies, the two quantities are related by

\begin{align}
& \nu(r)=\int \phi(M_\star,r)\,{\rm d}M_\star\,, \\
&\rho_{\rm gal}(r)=\int M_\star\phi(M_\star,r)\,{\rm d}M_\star
=\nu(r)\langle M_\star\rangle_r \,.
\end{align}
The radial shapes of $\nu(r)$ and $\rho_{\rm gal}(r)$ coincide only if
the mean stellar mass of the selected galaxies $\langle M_*\rangle_r$ and the sample
completeness are independent of the cluster-centric radius. This condition
might not hold because of mass segregation, dynamical friction, tidal
stripping, radial variations in the galaxy population, and the survey
selection function. Thus, the number density and the stellar mass profiles can both be approximately described by NFW models but
have different scale radii
\citep[e.g.][]{Annunziatella14}.

In the inference described in the next Section, we adopt wide Gaussian priors on the free parameters $\rho_{\rm g},\rho_{\star},r_{\rm g},r_{\star}$ obtained as follows. 
 
 For all the clusters, we consider the X-ray luminosity within $r_{500}$ from the meta-catalogue of X-ray detected clusters of galaxies (MCXC, \citealt{2011A&A...534A.109P}). We then infer the corresponding gas mass within $r_{500}$, i.e. $M_{\rm gas,500}\equiv M_{\rm gas}(<r_{500})$, using the empirical relations between the X-ray luminosity and the gas mass studied by  \cite{2022A&A...661A...7B}. In their study, the authors found  the following scaling relation between gas mass and X-ray luminosity:
\begin{equation}
    L_X=A\cdot L_{X,p}\left(\frac{M_{\rm gas}}{M_{{\rm gas},p}}\right)^B\left(\frac{E(z)}{E(z_p)}\right)^C\,,
    \label{Relation LX Mgas}
\end{equation}
where $A=0.89\pm0.02$, $B=1.10\pm0.03$, $C=1.44\pm0.26$ and $E(z)={H(z)}/{H_0}$.
The parameters $L_{X,p}=3.20\times10^{43}\,\text{erg}\ \text{s}^{-1}$, $ M_{{\rm gas},p}=1.04\times10^{13}\, \text{M}_{\odot} $, and $ z_p= 0.33 $ are the scale values of the sample in the study of  \cite{2022A&A...661A...7B}.

We inverted Eq.~\eqref{Relation LX Mgas} to estimate $M_{\rm gas,500}$ and the corresponding radius $r_{500}$ for each cluster. These two quantities fix the gas mass enclosed within $r_{500}$, but do not determine its radial distribution. To calibrate the profile shape, we used the few (three to five) clusters in each bin for which a resolved electron-density profile is available in the ACCEPT Chandra archive \citep{2009ApJS..182...12C}. We fitted each of these profiles independently with a $\beta$ model, this time allowing all the parameters  $\rho_{\rm g}$, $r_{\rm g}$, and $\beta$ to vary. We then computed the mean values of $\rho_{\rm g}$ and $\beta$ from the available fits in each bin, together with their cluster-to-cluster standard deviations. For every cluster $i$ in that bin, values of $\rho_{\rm g}^{(i)}$ and $\beta^{(i)}$ were drawn from Gaussian distributions centred on these mean values and with widths given by the corresponding average standard deviations.

Once $\rho_{\rm g}^{(i)}$ and $\beta^{(i)}$ had been assigned, the normalization $r_{\rm g}^{(i)}$ of each cluster was determined separately by requiring the reconstructed profile to reproduce its inferred gas mass within $r_{500}$,
\begin{equation}
M_{\rm gas,500}^{(i)}= 4\pi\rho_{\rm g}^{(i)}
\int_0^{r_{500}}{
\left[1+\left(\frac{r}{r_{\rm g}^{(i)}}\right)^2\right]^{-3\beta^{(i)}/2}
r^2\,{\rm d}r}.
\label{eq}
\end{equation}
The triplet $(M_{\rm gas,500}^{(i)},r_{\rm g}^{(i)},\beta^{(i)})$  therefore uniquely specifies the cumulative gas-mass profiles $M_{\rm gas}(<r)$ of every cluster.

\begin{figure}
\centering
\includegraphics[width=0.8\columnwidth]{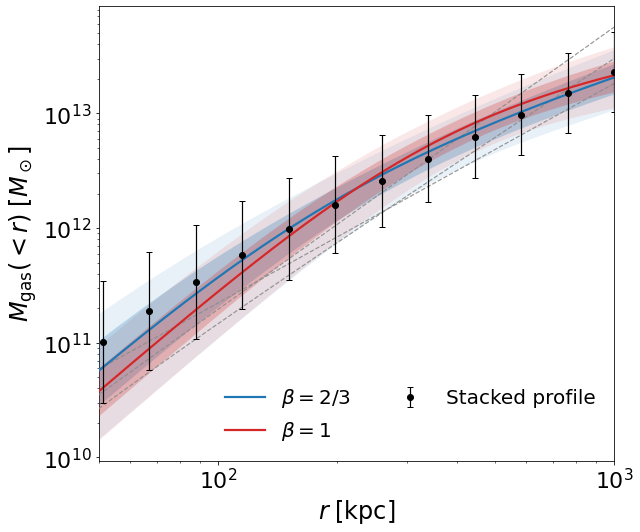}
\caption{Stacked gas mass profile (points with errorbars) for the clusters in Bin 1, obtained with the method described in Sect.~\ref{sec:gas}. The red and blue curves represent the best-fit $\beta = 1$ and $\beta = 2/3$-profiles, respectively, while the dark and light shaded bands indicate the corresponding one and two sigma uncertainties. Dashed lines show the reconstructed profiles from the electron density distributions of three clusters in Bin 1 found in the ACCEPT catalog.}
\label{Fig:gas profiles}
\end{figure}

Finally, within each cluster bin, we evaluated all the reconstructed gas-mass profiles on a common radial grid and stacked them to obtain the mean profile. At each radius, the uncertainty assigned to the stacked profile was taken to be the standard deviation among the individual cluster profiles, thereby retaining the cluster-to-cluster scatter rather than only the statistical uncertainty on the mean. The last step consisted in a final Monte Carlo-Markov-Chain (MCMC) fit of each stacked profile with the cumulative beta-model mass profile, considering separately the two fixed values $\beta=2/3$ and $\beta=1$, as mentioned above. We adopted wide, un-informative flat priors on $\rho_{\rm g},r_{\rm g}$; the covariance matrix $C(\rho_{\rm g},r_{\rm g})$ extracted from the posterior distribution is then used to impose the final Gaussian prior in our reference analysis. In Fig. \ref{Fig:gas profiles} we illustrate this procedure for Bin 1. The points with errobars indicate the stacked profile with the 68\% scatter, fitted with a $\beta$-model (red for $\beta =1$, blue for $\beta = 2/3$). The dashed profiles are the (extrapolated) gas mass distributions obtained for the three clusters in Bin 1 for which electron density data is present in the ACCEPT archive.

As for the stellar mass profile in galaxies, we considered the observations of the two massive galaxy clusters analysed in \cite{2025A&A...698A..83P}, AS1063 and MACS 1206, which include both gas and galaxies' radial mass profiles. We computed the fraction of galaxy mass with respect to gas $f=M_{\rm gal}/M_{\rm gas}$ and studied the evolution of $f$ as a function of radius for both clusters. We noticed that in both cases the fraction presents a rather similar profile with $f(r) \simeq 0.1$  for $r > 0.5$ Mpc.
As done in e.g. \cite{Famaey2025}, we assumed that the galaxy-to-gas mass fraction follows the same ratio across all the clusters in our sample and used it to derive the mean galaxy profile from the mean gas profile in each bin.  At this point, we fit the mean mass profile in galaxies assuming an NFW profile, 
using the same procedure employed for the gas mass, and considering the corresponding covariance matrix to obtain the Gaussian priors on the free parameters, namely the characteristic density $\rho_\star$ and the scale radius $r_\star$.
The median values of the posterior distributions of the galaxies and gas mass profile parameters with $\beta=2/3$ and $\beta=1$ are reported in Table~ \ref{tab:gas_galaxy_profiles} with their $1\sigma$ uncertainties.  Note that, in order to avoid over-constraining the reference analysis, we have conservatively broadened the resulting Gaussian priors by rescaling each posterior covariance matrix as $C\rightarrow 1.5^2 C$. 

%=============
\begin{table*}
\centering
\small
\begin{tabular}{c cc cc cc}
\toprule\\[-8pt]
& \multicolumn{2}{c}{$\beta=2/3$}
& \multicolumn{2}{c}{$\beta=1$}
& \multicolumn{2}{c}{mass profile in galaxies} \\[1ex]
Bin
& $\rho_{\rm g}\,[\text{M}_{\odot}/{\text{Mpc}^3}] $
& $r_{\rm g}$ [Mpc]
& $\rho_{\rm g}\,[{\text{M}_{\odot}}/{\text{Mpc}^3}] $ 
& $r_{\rm g}$ [Mpc]
& $\rho_\star\,[{\text{M}_{\odot}}/{\text{Mpc}^3}] $
& $r_\star$ [Mpc]\\
\midrule\noalign{\vskip 4pt}
Bin 1
& $\left(1.2^{+1.2}_{-0.6}\right)\times10^{14}$
& $0.13^{+0.08}_{-0.05}$
& $\left(7.4^{+4.6}_{-2.9}\right)\times10^{13}$
& $0.28^{+0.09}_{-0.07}$
& $\left(1.7^{+4.1}_{-1.2}\right)\times10^{12}$
& $0.59^{+0.62}_{-0.30}$ \\

Bin 2
& $\left(1.8^{+1.3}_{-0.8}\right)\times10^{14}$
& $0.14^{+0.06}_{-0.04}$
& $\left(1.1^{+0.5}_{-0.3}\right)\times10^{14}$
& $0.30^{+0.07}_{-0.06}$
& $\left(3.5^{+5.9}_{-2.2}\right)\times10^{12}$
& $0.53^{+0.42}_{-0.23}$ \\

Bin 3
& $\left(3.7^{+4.7}_{-2.1}\right)\times10^{14}$
& $0.12^{+0.08}_{-0.05}$
& $\left(2.0^{+1.3}_{-0.8}\right)\times10^{14}$
& $0.27^{+0.09}_{-0.06}$
& $\left(6.3^{+16.1}_{-4.6}\right)\times10^{12}$
& $0.48^{+0.52}_{-0.25}$ \\

Bin 1 relaxed
& $\left(1.1^{+1.1}_{-0.6}\right)\times10^{14}$
& $0.14^{+0.08}_{-0.05}$
& $\left(6.9^{+4.2}_{-2.6}\right)\times10^{13}$
& $0.30^{+0.09}_{-0.07}$
& $\left(1.7^{+4.2}_{-1.2}\right)\times10^{12}$
& $0.60^{+0.65}_{-0.31}$ \\

Bin 2 relaxed
& $\left(2.8^{+3.1}_{-1.5}\right)\times10^{14}$
& $0.12^{+0.07}_{-0.04}$
& $\left(1.5^{+0.9}_{-0.5}\right)\times10^{14}$
& $0.29^{+0.08}_{-0.06}$
& $\left(4.1^{+7.0}_{-2.6}\right)\times10^{12}$
& $0.52^{+0.42}_{-0.23}$ \\
\bottomrule
\end{tabular}
\caption{Best-fit parameters of the gas profiles modelled with a $\beta$ model for $\beta=2/3$ (columns two and three) and $\beta=1$ (columns four and five), together with the parameters of the galaxy mass profiles modelled with an NFW profile.}
\label{tab:gas_galaxy_profiles}
\end{table*}

\section{Kinematical analysis with \texttt{MG-MAMPOSSt}}
\label{sec:mgmamposst}
We reconstruct the total gravitational potential of the stacked clusters in both Newtonian and RG frameworks with the \texttt{MG-MAMPOSSt} code \citep{2021MNRAS.506..595P, Pizzuti2023JOSS}, an extension of the \texttt{MAMPOSSt} method \citep{2013MNRAS.429.3079M} specifically designed for kinematical studies of galaxy clusters in general MG models. The original \texttt{MAMPOSSt} algorithm performs a maximum-likelihood fit of the p.p.s distribution of cluster members $(R,v_z)$ and, under the assumptions of spherical symmetry and dynamical relaxation, it jointly constrains the gravitational potential and the orbital anisotropy of the member galaxy population. The method is based on the spherical Jeans equation,
\begin{equation}
    \frac{d}{dr}\left[\nu(r)\sigma_r^2(r)\right]
    +2\beta(r)\frac{\nu(r)\sigma_r^2(r)}{r}
    =
    -\nu(r)\frac{d\Phi}{dr},
    \label{eq:jeans_mg}
\end{equation}
where $\nu(r)$ is the three-dimensional number-density profile of the tracer galaxies, $\sigma_r(r)$ is their radial velocity-dispersion profile, and $\Phi(r)$ is the total gravitational potential. The velocity anisotropy profile is defined as
\begin{equation}
    \beta(r)
    \equiv
    1-\frac{\sigma_\theta^2(r)+\sigma_\phi^2(r)}
    {2\sigma_r^2(r)} ,
    \label{eq:beta_def}
\end{equation}
with $\beta=0$ corresponding to isotropic orbits, $\beta>0$ to radially biased orbits, and $\beta<0$ to tangentially biased orbits. For a given choice of the mass profile, tracer number-density profile, and anisotropy model, \texttt{MAMPOSSt} solves Eq.~\eqref{eq:jeans_mg}, projects the resulting velocity distribution along the line of sight, and compares the predicted distribution directly with the observed p.p.s of the cluster. The current implementation of the method assumes a Gaussian ansatz for the three-dimensional velocity distribution and negligible streaming motions.
\texttt{MG-MAMPOSSt}  generalizes this framework by replacing the standard Newtonian acceleration entering the Jeans equation with general parametrization of the gravitational potential predicted in a wide variety of models. In the limit of Newtonian gravity, the code reduces to the standard \texttt{MAMPOSSt} analysis, whereas in MG models the potential includes the additional force contribution associated with the modified dynamics; for the specific case of RG, the expression is given by Eq.~\eqref{Gradient Phi}.

The \texttt{MAMPOSSt} method has been tested extensively on mock galaxy catalogues extracted from numerical simulations, showing that it can recover cluster mass profiles and anisotropy parameters with small bias and controlled scatter, even in the presence of moderate deviations from the Gaussian velocity ansatz \citep[e.g.][]{2013MNRAS.429.3079M,Old_2015,Read2021}. Both \texttt{MAMPOSSt} and \texttt{MG-MAMPOSSt} have been widely applied to perform detailed kinematical analyses of individual and stacked galaxy clusters \citep[e.g.][]{Biviano_2013,Capasso_2019,Mamon2019,Biviano_2023,Pizzuti26}.

For the the anisotropy profile, here  we adopt a parametric ansatz based on the generalized Tiret model (gT, hereafter), 
\begin{equation}
    \beta(r)=\beta_0+\frac{(\beta_{\infty}-\beta_0)r}{r+r_{\beta}} \,,
    \label{anisotropy gT}
\end{equation}
which extends the Tiret profile of \cite{2007A&A...476L...1T}. In the equation above, $r_{\beta}$ is a characteristic scale radius and $\beta_{\infty}$ and $\beta_0$ are the values of the anisotropy respectively at infinity and at $r=0$. This model is rather general and it captures a wide variety of galaxies' orbits \cite[e.g.,][]{Mamon2019,2025A&A...699A..88P}. In the fit, as typically done in similar works, we consider the parameters $\mathcal{A}_{0/\infty}=(1-\beta_{0,\infty})^{-1/2}$, which are strictly positive and larger than one for radial orbits.
 As for the scale radius, preliminary tests showed that no improvement in the posterior arises if $r_\beta$ is treated as a free parameter, and its distribution is basically un-constrained. As such, in the following we fix $r_\beta = 0.3$ Mpc, which is close to the best fit values for all the bins.

We perform our analysis in two steps.  First, we derive the mass and anisotropy profile of our stacked system in Newtonian gravity, which will serve as a baseline for comparison of the results in RG.  We applied the \texttt{MG-MAMPOSSt} code to infer the total (dark+baryonic) mass profile of each system by assuming a NFW model for the DM component, while the mass profiles of gas and galaxies are modelled as described above. 
Afterwards, we run again \texttt{MG-MAMPOSSt} for the phenomenological RG framework with no DM to reconstruct the gravitational potential and the average baryonic mass, and to constrain the vacuum gravitational permittivity.

In both gravity scenarios, we consider only galaxies with projected radius $ R\geq 0.05$ Mpc and $ R\leq1.2\times \bar{r}_{200}$; as discussed in Sect.~\ref{sec:RG}, the central region is excluded since the dynamics is dominated by the BCG. As for the upper bound, galaxies in the outskirts of clusters are no longer expected to be in dynamical equilibrium and may belong to the surrounding large-scale structure (see, e.g.,~\citealt{Pedratti26}). We tested that variations of up to 10\% in the upper limit do not produce significant changes in the final results.

We perform a MCMC exploration of the parameter space with 110000 points, ensuring convergence by requiring that the Gelman-Rubin coefficient $\hat{R}$ remains always less than 1.1. For the Newtonian run, we consider five free parameters to be optimized, $r_{200}$ and the scale radius $r_s$ for the NFW dark matter profile, $\mathcal{A_{\infty}}$ and $\mathcal{A}_0$ for the anisotropy model, and the scale radius of the tracer density profile, $r_{\nu}$.
For the latter, we assume that $\nu(r)$ follows a NFW profile, which has been shown to provide a good fit to the distribution of member galaxies in clusters (see e.g. \citealt{2004ApJ...610..745L, 2012MNRAS.427..428G, 2012MNRAS.423..104B,Annunziatella14, 2017A&A...606A.108C}). Note that, as mentioned above, the numerical profile (and its slope) does not necessarily coincide with the stellar mass in galaxies. As such, we will consider the scale radii $r_\nu$ and $r_\star$ as independent free parameters to be marginalised in our inference.

In principle, $r_\nu$ can be estimated directly from the p.p.s before running the \texttt{MG-MAMPOSSt} procedure. As done in several works \citep[e.g.,][]{Biviano_2023,2025A&A...698A..83P},  we perform a maximum likelihood fit of the projected NFW profile to the distribution of projected radii of member galaxies in each stacked configuration. We applied the method of \cite{Sarazin_1980} which does not require the binning of data \citep[e.g.,][]{Biviano_2013}. The fit is performed over the same
radial interval and using the same final tracer sample adopted in the \texttt{MG-MAMPOSSt} analysis.  
Note that homogeneous estimates of the radial spectroscopic completeness are not available for the individual CIRS clusters; the analyses of \citet{2006AJ....132.1275R} and 
\citet{Pizzardo_2021} indicate an average spectroscopic completeness of $\gtrsim$80\% within $R_{200}$, but  no information can  be found on its radial profile within the virial radius. We therefore cannot apply an explicit completeness correction to the stacked radial distribution. Stacking increases the number of tracers but does not, by itself, remove a possible radial dependence of the spectroscopic selection function. The inferred $r_\nu$ should consequently be interpreted as the effective scale radius of the observed tracer population in each stack. If the residual incompleteness is approximately independent of projected radius, it affects only the
normalization of the number density profile which does not enter the Jeans analysis; a radial dependence could instead bias the inferred profile shape.

In the \texttt{MG-MAMPOSSt} fit, $r_\nu$ is then allowed to vary with a flat prior whose bounds are set by the 95\% region of the external fit. We get $r_\nu \in [0.30,0.50], [0.36,0.52], [0.58,0.75], [0.28, 0.48]$, $ [0.37,0.51]\,\text{Mpc}$, for Bin 1,2,3, relaxed 1 and relaxed 2, respectively. For all the other parameters, we consider flat uninformative priors $r_{200} \in [0.5, 5.0]$ Mpc, $r_\text{s}\in[0.04, 4.0]$ Mpc,  $\mathcal{A}_{0,\infty}\in[0.4,5.0]$.  The four parameters describing the gas and galaxy profiles are also optimized in the MCMC with the Gaussian priors discussed in Sect.~\ref{sec:gas}.

For the RG analysis, the two parameters describing the NFW mass profile are replaced by the three parameters entering the gravitational-permittivity function $\epsilon(\rho)$ defined in Eq.~\ref{epsilon (rho)}: the vacuum value $\epsilon_0$, the critical density $\rho_c$, and the transition steepness $Q$. We assume the following broad uniform priors:
\begin{equation}
\begin{split}
& \epsilon_0\in[0.001,0.999],\,\,
\log_{10}\left[
\frac{\rho_c}{\text{M}_\odot\,\mathrm{Mpc}^{-3}}
\right]\in[10,20],\\
& Q\in[0.01,11].
\end{split}
\end{equation}
All remaining model parameters and their priors are kept identical to those adopted in the Newtonian analysis. 

\section{Results}
\label{sec:results}

\subsection{Newtonian Gravity}
The line-of-sight velocity dispersion profiles reconstructed with the \texttt{MG-MAMPOSSt} analysis are shown in Figs.~\ref{fig:VDP profile} (main bins) and \ref{fig:VDP profile relaxed Bin} ("relaxed" bins) as the blue curves and shaded regions (68\% confidence level). The upper plots of both Figures are for the case of $\beta = 2/3$ in the gas profile, while the bottom plots indicate the results for $\beta = 1$. The best fit values of all the parameters for each bin are further reported in Table \ref{tab: pars GR}.
\begin{table*}
{\small
\setlength{\tabcolsep}{4pt}
\renewcommand{\arraystretch}{0.95}
    \centering
    
    \begin{tabular}{cccccc}
    \toprule
    $\beta = 2/3$  & $r_{200}$ [Mpc] & $r_{\nu}$ [Mpc]	 & $r_{s}$ [Mpc]	 & $\mathcal{A}_\infty  $	 & $\mathcal{A}_0$   \\[1ex]
        \hline\noalign{\vskip 4pt}
      Bin 1 & $0.61^{+0.23}_{-0.07}$ & $0.45^{+0.04}_{-0.08}$ & $0.08^{+0.31}_{-0.02}$ & $1.5^{+1.8}_{-0.6}$ & $0.54^{+0.64}_{-0.14}$ \\[1ex]
  
      Bin 2 & $0.95^{+0.11}_{-0.09}$ & $0.40^{+0.06}_{-0.02}$ & $0.17^{+0.13}_{-0.01}$ & $2.1^{+1.8}_{-0.8}$ & $0.64^{+0.27}_{-0.10}$ \\[1ex]
    
       Bin 3 & $1.15^{+0.27}_{-0.14}$ & $0.65^{+0.06}_{-0.05}$ & $0.22^{+0.27}_{-0.06}$ & $1.6^{+1.8}_{-0.5}$ & $0.75^{+0.35}_{-0.21}$ \\[1ex]
     
       Bin 1 relaxed & $0.76^{+0.05}_{-0.06}$ & $0.35^{+0.06}_{-0.05}$ & $0.20^{+0.11}_{-0.12}$ & $3.3^{+1.3}_{-2.0}$ & $0.65^{+0.30}_{-0.25}$ \\[1ex]
     
      Bin 2 relaxed & $1.19^{+0.02}_{-0.30}$ & $0.42^{+0.08}_{-0.03}$ & $0.29^{+0.03}_{-0.22}$ & $1.7^{+1.6}_{-0.6}$ & $0.86^{+0.23}_{-0.46}$  \\[1ex]
       \hline\noalign{\vskip 4pt}
        $\beta = 1$  &  $r_{200}$ [Mpc] & $r_\nu$  [Mpc] &  $r_{s}$ [Mpc] & $\mathcal{A}_\infty  $  & $\mathcal{A}_0$  \\[1ex]
        \hline\noalign{\vskip 4pt}
         Bin 1 &$0.82^{+0.06}_{-0.21}$ & $0.43^{+0.06}_{-0.07}$ & $0.34^{+0.14}_{-0.27}$ & $1.8^{+2.0}_{-0.8}$ & $0.84^{+0.36}_{-0.44}$\\[1ex]
        Bin 2 & $0.85^{+0.20}_{-0.09}$ & $0.40^{+0.07}_{-0.02}$ & $0.16^{+0.18}_{-0.00}$ & $2.0^{+1.9}_{-0.7}$ & $0.68^{+0.30}_{-0.13}$\\[1ex]
        Bin 3 & $1.31^{+0.15}_{-0.58}$ & $0.61^{+0.09}_{-0.01}$ & $0.36^{+0.21}_{-0.21}$ & $1.8^{+1.2}_{-0.7}$ & $0.89^{+0.36}_{-0.30}$ \\[1ex]
         Bin 1 relaxed & $0.72^{+0.07}_{-0.13}$ & $0.35^{+0.05}_{-0.05}$ & $0.17^{+0.13}_{-0.10}$ & $3.2^{+1.3}_{-2.0}$ & $0.60^{+0.38}_{-0.20}$ \\[1ex]
         Bin 2 relaxed & $1.18^{+0.05}_{-0.29}$ & $0.43^{+0.06}_{-0.05}$ & $0.29^{+0.09}_{-0.22}$ & $1.9^{+1.5}_{-0.8}$ & $0.82^{+0.33}_{-0.41}$ \\
    \bottomrule
    \end{tabular}
    \caption{Best fit values for the free parameters in the run in Newtonian gravity. The DM profile is modelled with an NFW profile, while the gas and galaxy profiles are added separately. The uncertainties reported are all at the 95\% confidence level.}
    \label{tab: pars GR}}
\end{table*}
We checked that the constraints obtained for the $\beta= 2/3$ and $\beta=1$ cases are in good agreement within the uncertainties, and provide a negligible difference in the $\chi^2$, demonstrating that the systematics induced by the choice of the ICM parametrisation in Newtonian gravity are subdominant compared to statistical uncertainties.

Interestingly, the values of $r_{200}$ inferred from the stacked p.p.s do not coincide exactly with the mean values $\bar r_{200}$ reported in Table~\ref{tab:Binning}. The two quantities, however, are not equivalent estimators: the tabulated $\bar r_{200}$ values are averages of the individual caustic-based measurements, whereas \texttt{MG-MAMPOSSt} determines the effective dynamical scale of a single model fitted to the full stacked galaxy distribution. Since each galaxy contributes separately to the likelihood, clusters are implicitly weighted by the number and phase-space distribution of the members retained after the \texttt{Clean} selection. Moreover, the relation between $r_{200}$ and the projected velocity distribution is non-linear and depends jointly on the kinematical quantities (tracers distribution, velocity anisotropy, scale radius of the mass profile) in the Jeans' equation. The fitted $r_{200}$ should therefore be interpreted as the characteristic dynamical radius of the ensemble cluster, which is not a direct reconstruction of the mean $\bar r_{200}$.

\subsection{Refracted Gravity}
\begin{figure*}
\centering
\includegraphics[width=0.73\textwidth]{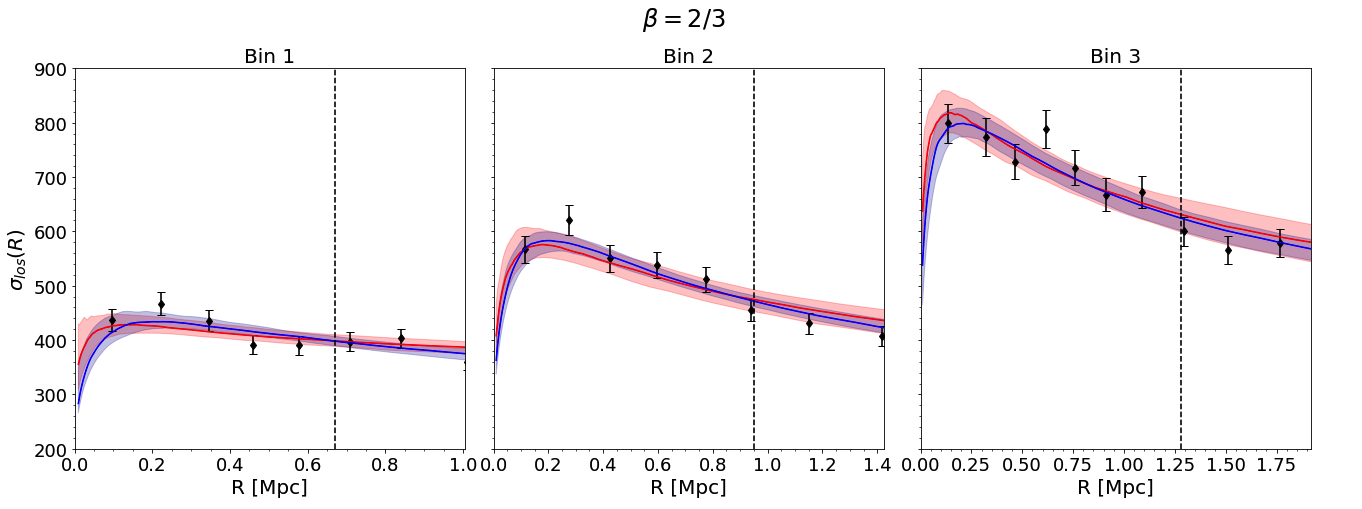}
\includegraphics[width=0.73\textwidth]{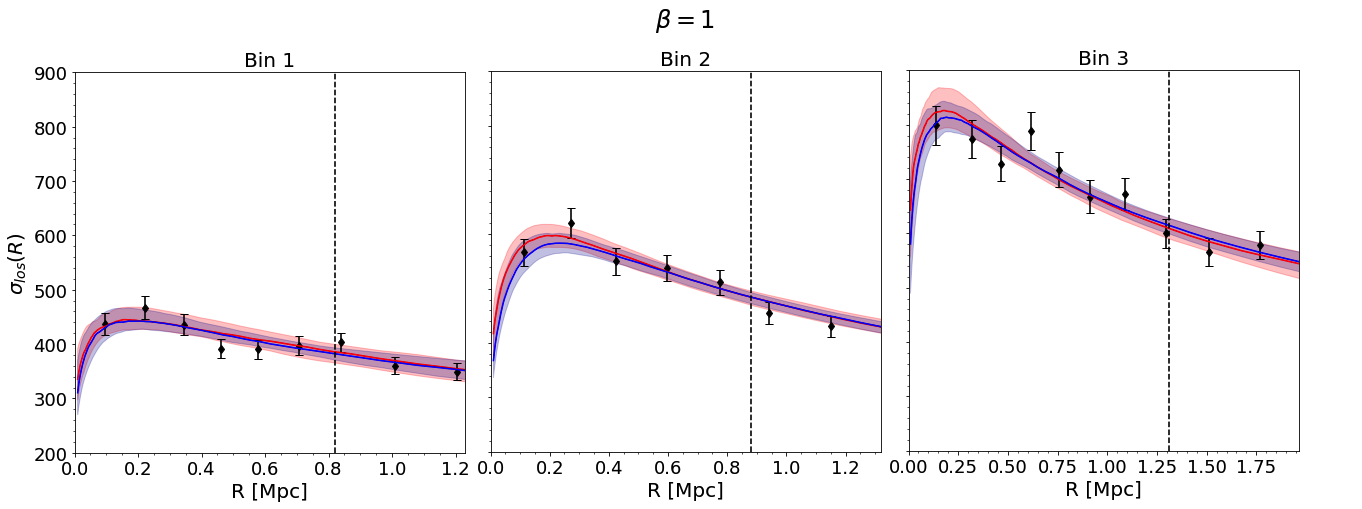}
\caption{Line-of-sight velocity dispersion profiles of the stacked clusters (black points) fitted assuming Newtonian gravity with a NFW dark matter model (blue), and RG (red). The shaded regions correspond to $1\sigma$ uncertainties and the dashed lines indicate the best fit value of $r_{200}$ of the DM profile in each bin. The gas is modelled with a $\beta$-profile. Top: $\beta = 2/3$. Bottom: $\beta = 1$ }
\label{fig:VDP profile}
\end{figure*}

\begin{figure}
\centering
\includegraphics[width=\columnwidth]{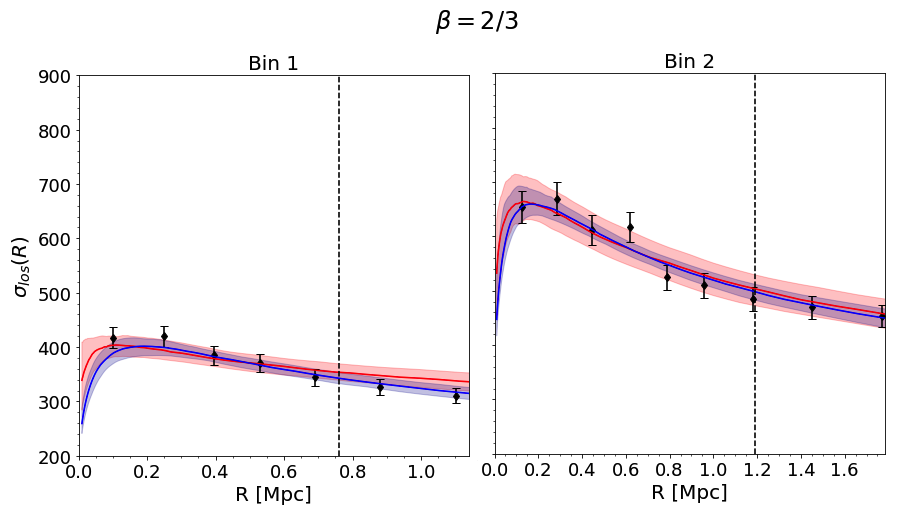}
\includegraphics[width=\columnwidth]{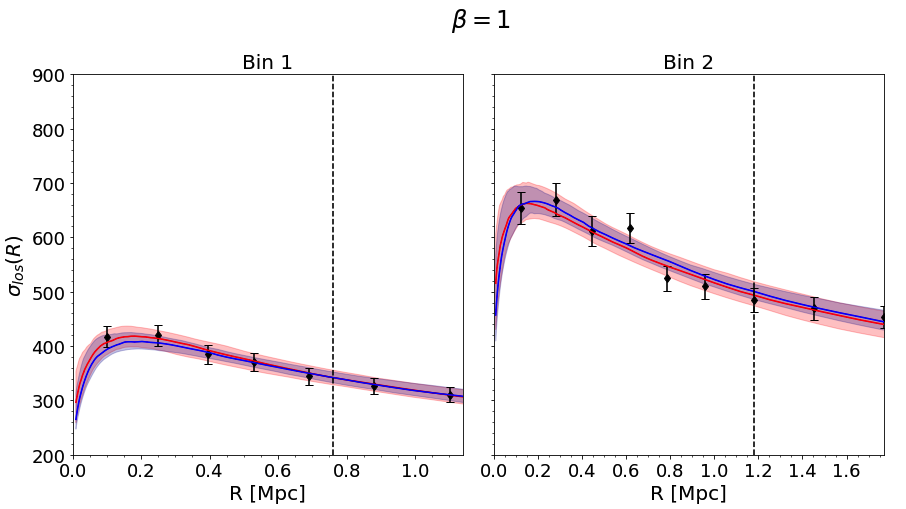}
\caption{Same as Fig.~\ref{fig:VDP profile}, but considering the two bins with "relaxed" galaxy clusters only (i.e., $A^2 < 0.7$).}
\label{fig:VDP profile relaxed Bin}
\end{figure}

The red curves and shaded regions in Figs.~\ref{fig:VDP profile} and \ref{fig:VDP profile relaxed Bin} show the line-of-sight velocity dispersion obtained by the \texttt{MG-MAMPOSSt} analysis in RG. The constraints on the free parameters are listed in Tab.~\ref{tab:RG parameters} for the $\beta=2/3$ case in the first five rows, and for the $\beta=1$ case in the last five rows. As for the Newtonian run, we found an overall good agreement among the results within uncertainties. Moreover, the common kinematical parameters ($\mathcal{A_0},\mathcal{A}_\infty, r_\nu$) between the two gravitational frameworks show similar bounds, differently from what was found for the velocity anisotropy profile in \cite{2025A&A...698A..83P}.

The constraints on the RG parameters are obtained after
marginalizing over the gas and stellar mass-profile parameters.
This approach propagates the uncertainty in the adopted baryonic
description into the inferred gravitational permittivity.
For the three main stacks, the central estimates of
$\epsilon_0$ range from $0.04$ to $0.08$ for $\beta=2/3$
and from $0.19$ to $0.37$ for $\beta=1$. Although the constraints remain compatible within the quoted
uncertainties, the case $\beta=1$ indicates a preference for larger values of $\epsilon_0$, particularly evident for Bin 3; these shifts indicate an appreciable sensitivity
to the assumed gas-density slope.

The two stacks selected with $A^2<0.7$ provide a more consistent
comparison across the two gas models. Their central estimates
are $\epsilon_0=0.04$ and $0.06$ for $\beta=2/3$, and
$\epsilon_0=0.04$ and $0.05$ for $\beta=1$.
The agreement between the two selected stacks is therefore
maintained for both gas parametrizations, with smaller shifts
in the central estimates when the gas-density slope is changed.
This provides an internal consistency result, within the adopted
modelling framework, that a dynamical-state-based selection may help in reducing the systematics uncertainties. 
 
As for the other two RG parameters,  the large uncertainties in the distribution prevent us from obtaining accurate constraints. However, while the transition exponent $Q$ is totally unbound, we can still pose a lower limit on the critical density, $\log_{10}[\rho_c/ (\text{M}_\odot/\text{Mpc}^3)] \gtrsim 12.3$, consistent in all bins. We also note that the RG parameters show no relevant correlations among each other and with the kinematical parameters of anisotropy and number density profiles. A slight positive correlation is present between the vacuum permittivity and the baryonic profile parameters, as shown in the example of Appendix \ref{app:A}.
\begin{table*}
    \centering
    {\small
\setlength{\tabcolsep}{4pt}
\renewcommand{\arraystretch}{0.95}
    \begin{tabular}{ccccccc}
    \toprule
        $\beta = 2/3$  & $r_\nu$  [Mpc]&  $\mathcal{A}_\infty  $  & $\mathcal{A}_0$  & $\epsilon_0$ & $\log_{10}[\rho_c/(\text{M}_{\odot}/\text{Mpc}^3)]$ & $Q$ \\
        \hline\noalign{\vskip 4pt}
        Bin 1 & $ {0.44}\pm 0.06$ & $1.5^{+0.9}_{-0.5}$ & $1.00^{+0.65}_{-0.21}$ & $0.08^{+0.26}_{-0.06}$ & $\geq13.09$& - \\[1ex]
        
        Bin 2& ${0.42}_{-0.04}^{+0.07}$ &$2.1^{+1.4}_{-0.8}$ & $0.84^{+0.25}_{-0.13}$ & $0.04^{+0.20}_{-0.02}$& $\geq 14.25$ & -\\[1ex]
        
        Bin 3& ${0.65}_{-0.05}^{+0.09}$ & $1.9^{+1.7}_{-0.8}$ & $0.93^{+0.31}_{-0.19}$  & $0.05^{+0.38}_{-0.04}$ & $\geq 13.90$& -\\[1ex]
        
        Bin 1 relaxed & ${0.37}_{-0.06}^{+0.07}$ &$1.6^{+2.0}_{-0.7}$& $1.02^{+0.61}_{-0.24}$& $0.04^{+0.21}_{-0.03}$ & $\geq12.76$ & - \\[1ex]
        
        Bin 2 relaxed & $0.44^{+0.07}_{-0.05}$ & $1.5^{+1.4}_{-0.4}$ & $0.93^{+0.40}_{-0.24}$ &${0.06}_{-0.03}^{+0.32}$ &$\geq 13.99$ &- \\[1ex]
        \hline\noalign{\vskip 4pt}
        $\beta = 1$  & $r_\nu$  [Mpc] &  $\mathcal{A}_\infty  $  & $\mathcal{A}_0$  & $\epsilon_0$ & $\log_{10}[\rho_c/(\text{M}_{\odot}/\text{Mpc}^3)]$ & $Q$ \\[1ex]
        \hline\noalign{\vskip 4pt}
        Bin 1 & $0.42^{+0.07}_{-0.05}$ & $1.4^{+1.7}_{-0.4}$ & $1.01^{+0.48}_{-0.29}$ &  $0.20^{+0.30}_{-0.15}$ & $\geq 14.37$&- \\[1ex]
        
        Bin 2& ${0.41}_{-0.03}^{+0.07}$ & $1.8^{+1.1}_{-0.6}$ & $0.97^{+0.28}_{-0.17}$ & $0.19^{+0.38}_{-0.07}$ & $\geq 12.33$ &-\\[1ex]
        
        Bin 3& ${0.64}_{-0.04}^{+0.08}$ & $1.6^{+1.4}_{-0.5}$ & $1.02^{+0.44}_{-0.24}$ & $0.37^{+0.37}_{-0.29}$ & $\geq 14.72$& -\\[1ex]
        
        Bin 1 relaxed & $0.34^{+0.08}_{-0.04}$ & $1.9^{+2.4}_{-0.7}$ & $0.93^{+0.44}_{-0.19}$& $0.04^{+0.22}_{-0.03}$ & $\geq 13.19$ & - \\[1ex]
        
        Bin 2 relaxed & $0.44^{+0.07}_{-0.05}$ & $1.6^{+1.2}_{-0.5}$ & $0.93^{+0.51}_{-0.22}$ & $0.05^{+0.19}_{-0.03}$ &$\geq 13.23$ &- \\[1ex]
        \bottomrule
    \end{tabular}
    \caption{Best fit values of the free parameters of the run in RG with the gas profile modelled with $\beta=2/3$ (first five rows) and $\beta = 1$ (last five rows). The uncertainties reported are all at the 95\% confidence level.}
    \label{tab:RG parameters}}
\end{table*}

\begin{figure}
    \centering
    \includegraphics[width=0.73\linewidth]{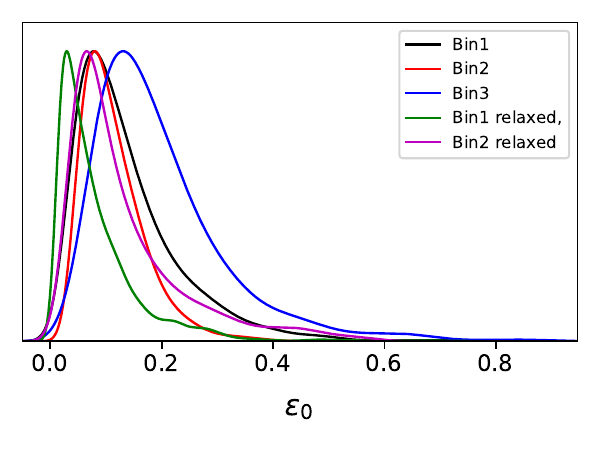}
    \includegraphics[width=0.73\linewidth]{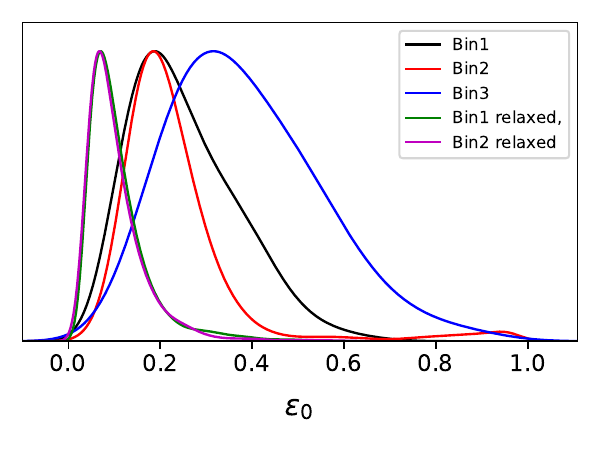}
    \caption{Comparison of the distribution of the vacuum permittivity $\epsilon_0$ estimated with the \textsc{MG-MAMPOSSt} code with the gas modelled with  with a $\beta$-profile. Top: $\beta = 2/3$. Bottom: $\beta = 1$.}
    \label{epsilon comparison}
\end{figure}

In order to assess which model better describes the data between RG and standard gravity,  we computed the BIC, which is defined as $\text{BIC} = -2\ln \mathcal{L}_{\max} +k\ln(N) $ where $\mathcal{L}_{\max}$ is the value of the likelihood at the best fit (i.e., the maximum), $k$ is the number of free parameters of the model, and $N$ is the number of data points. In our case, the analysis in RG has 10 free parameters (including the baryonic ones), while the one in the Newtonian framework considers 9 free parameters, and N is the total number of galaxies considered.  
We find that Newtonian gravity is significantly preferred with $\Delta \text{BIC}>6$ for all bins. However, this result is mostly determined by the large penalty given by the BIC to the model with more free parameters, as the values of the likelihood at the best fit do not present a significant difference.
The implications of these results are discussed in Sect.~\ref{sec:discussion}.

\subsection{Reconstructed gas profiles}

We finally examine the baryonic profiles inferred by \texttt{MG-MAMPOSSt} for the different stacked configurations. We focus on the ICM, which dominates the baryonic mass budget, and consider both fixed values of the beta-model exponent. The best-fitting parameters obtained in RG and Newtonian gravity are reported in Table~\ref{tab.gasfinalbeta_23} for $\beta=2/3$ (upper five rows) and for $\beta=1$ (lower five rows).

For $\beta=1$, the gas-mass profiles inferred in the two gravitational frameworks are consistent within $1\sigma$ over the radial range considered.  
When $\beta=2/3$ is adopted, both $\rho_{\rm g}$ and $r_{\rm g}$ tend to be slightly lower in RG than in Newtonian gravity. The resulting gas-mass profiles are lower in the cluster outskirts, although the differences remain within the $2\sigma$ uncertainties. This behaviour can be understood from the shallower outer density slope of the $\beta=2/3$ model, $\rho_{\rm gas}\propto r^{-2}$, for which the cumulative gas mass grows approximately linearly with radius. As a matter of fact, in RG -- where the gravitational field is sourced by the baryonic distribution alone -- this density slope produces a larger gravitational contribution in the outskirts. The inference partially compensates for this growth, favouring lower values of $\rho_{\rm g}$ and $r_{\rm g}$.

Consistently, changing $\beta$ produces no appreciable variation in the maximum likelihood of the Newtonian analysis, whereas the $\beta=1$ model provides a slightly better fit in RG. The different response of the reconstructed gas profiles highlights the sensitivity of RG to the assumed ICM distribution. 
Changing the gas-density slope produces appreciable shifts in the central estimates of $\epsilon_0$ for the three main stacks, although their quoted credible intervals overlap. Much smaller shifts are found for the two stacks selected with $A^2<0.7$.
The re-optimization of the baryonic parameters allows both gas prescriptions to describe the kinematical data, but does not establish the degree of uncertainty associated with ICM modelling. Indeed, the model dependence associated with these two ICM prescriptions is subdominant given the wide current statistical uncertainty. At the present level of precision, this result supports our use of a simple single-beta parametrization rather than a more flexible gas model with additional, poorly constrained parameters. This conclusion is specific to the present data and radial coverage; more precise analyses will require resolved observed ICM profiles and a broader assessment of the associated modelling uncertainty.

\begin{table*}
\centering
\small
\begin{tabular}{c cc cc}
\toprule
& \multicolumn{2}{c}{ RG}
& \multicolumn{2}{c}{ GR}\\

$\beta = 2/3$
& $\rho_{\rm g}\,[M_{\odot}/{\text{Mpc}^3}] $
& $r_{\rm g}$ [Mpc]
& $\rho_{\rm g}\,[{M_{\odot}}/{\text{Mpc}^3}] $ 
& $r_{\rm g}$ [Mpc]\\
\midrule
Bin 1
& $\left(1.7^{+1.8}_{-1.1}\right)\times 10^{14}$ & $0.07^{+0.05}_{-0.03}$
& $\left(1.9^{+1.3}_{-1.6}\right)\times 10^{14}$ & $0.13\pm 0.08$ \\[1ex]
Bin 2 &  $\left(1.9^{+1.4}_{-1.2}\right)\times10^{14}$ & $0.08^{+0.05}_{-0.03}$& $\left(1.5^{+2.4}_{-1.2}\right)\times10^{14}$
& $0.14^{+0.03}_{-0.09}$ \\[1ex]
Bin 3
& $( 2.3^{+4.5}_{-1.2})\times10^{14}$
& $0.04^{+0.06}_{-0.02}$ & $\left(6.6^{+5.1}_{-5.5}\right)\times10^{14}$ & $0.10^{+0.09}_{-0.08}$\\[1ex]
Bin 1 relaxed &
$\left(1.1^{+0.9}_{-0.8}\right)\times 10^{14}$ & $0.08^{+0.07}_{-0.03}$
& $\left(1.4^{+1.7}_{-1.0}\right)\times 10^{14}$ 
& $0.12^{+0.05}_{-0.07}$ \\[1ex]
Bin 2 relaxed
& $\left(1.7^{+4.7}_{-1.2}\right)\times 10^{14}$ & $0.06^{+0.08}_{-0.03}$
& $\left(5.6^{+12.8}_{-4.1}\right)\times 10^{13}$ & $0.11^{+0.23}_{-0.05}$\vspace{2pt} \\
\midrule
$\beta = 1$
& $\rho_{\rm g}\,[M_{\odot}/{\text{Mpc}^3}] $
& $r_{\rm g}$ [Mpc]
& $\rho_{\rm g}\,[{M_{\odot}}/{\text{Mpc}^3}] $ 
& $r_{\rm g}$ [Mpc]\\
\midrule
Bin 1
& $\left(9.8^{+6.2}_{-6.6}\right)\times 10^{13}$ & $0.19^{+0.15}_{-0.08}$ & $\left(7.3^{+12.2}_{-5.0}\right)\times 10^{13}$ & $0.24^{+0.21}_{-0.09}$ \\ [1ex]

Bin 2
& $\left(1.1^{+0.7}_{-0.6}\right)\times10^{14}$
& $0.29^{+0.09}_{-0.06}$
&$\left(9.6^{+5.7}_{-6.6}\right)\times10^{13}$
&$0.40^{+0.11}_{-0.17}$\\[1ex]

Bin 3
& $\left(2.3^{+1.9}_{-1.5}\right)\times 10^{14}$ & $0.28^{+0.12}_{-0.14}$  & $\left(2.2^{+2.1}_{-1.9}\right)\times 10^{14}$ & $0.29^{+0.13}_{-0.17}$ \\[1ex]

Bin 1 relaxed
& $\left(1.0^{+3.0}_{-0.8}\right)\times10^{14}$
& $0.22^{+0.13}_{-0.07}$
&$\left(1.0^{+2.5}_{-0.8}\right)\times10^{14}$
& $0.30^{+0.23}_{-0.13}$\\[1ex]

Bin 2 relaxed
& $\left(1.2^{+4.5}_{-0.9}\right)\times 10^{14}$ & $0.23^{+0.15}_{-0.06}$
& $\left(1.5^{+3.5}_{-1.1}\right)\times 10^{14}$ & $0.29^{+0.24}_{-0.12}$\\
\bottomrule
\end{tabular}
\caption{Best fit of the free parameters of the gas mass profile in RG and GR with $\beta=2/3$ (upper five rows) and $\beta = 1$ (lower five rows). The uncertainties indicates the 95\% confidence level.}
\label{tab.gasfinalbeta_23}
\end{table*}

\section{Discussion}
\label{sec:discussion}
\begin{figure}
    \centering
    \includegraphics[width=\linewidth]{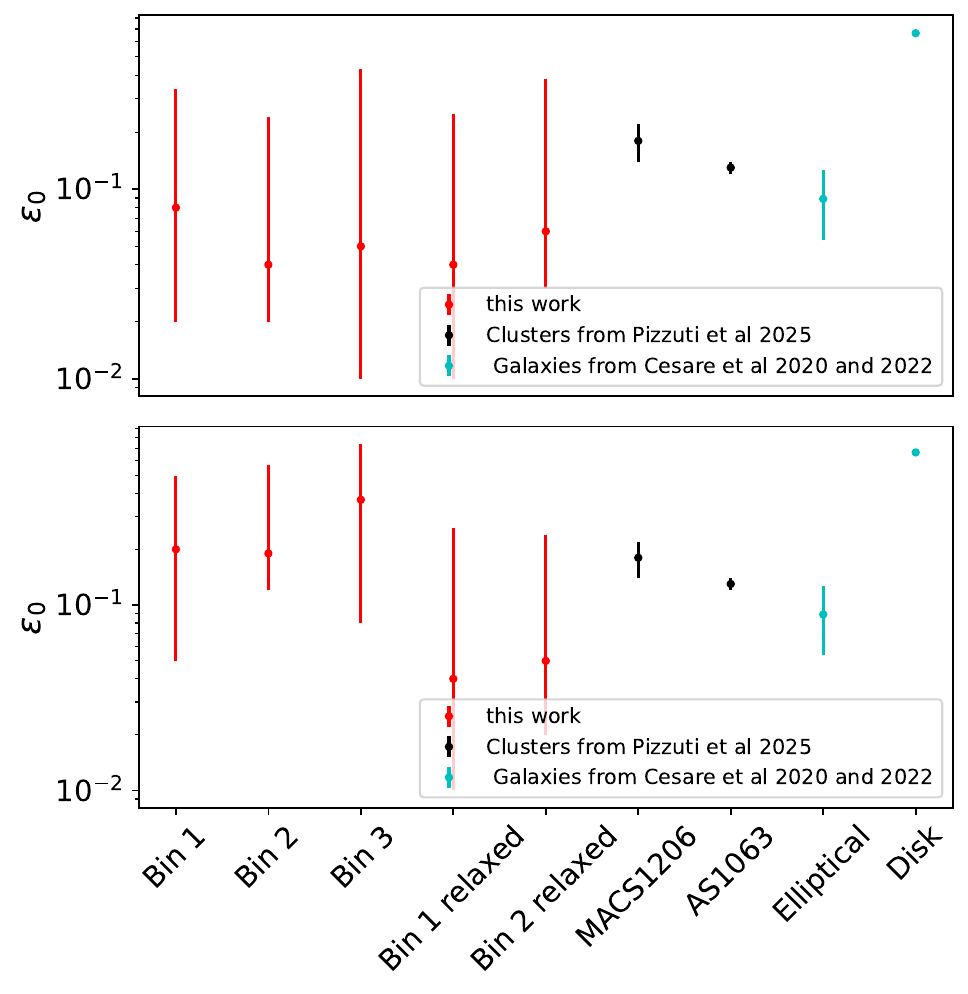}
    
    \caption{Comparison of the constraints on the vacuum gravitational permittivity $\epsilon_0$ obtained in this work with those derived from individual galaxy clusters \citep{2025A&A...698A..83P}, disk galaxies \citep{2020A&A...637A..70C}, and E0 elliptical galaxies \citep{2022A&A...657A.133C}. The upper and lower panels show the results obtained by modelling the ICM with $\beta=2/3$ and $\beta=1$, respectively.}
    \label{comparison epsilon}
\end{figure}

In this section, we compare our results on the vacuum permittivity with previous tests on RG both at cluster \citep{2025A&A...698A..83P}, and galactic  \citep{2020A&A...637A..70C,2022A&A...657A.133C} scales. The constraints are displayed in Fig.~\ref{comparison epsilon}. 

We found that the values of $\epsilon_0$ inferred from the stacked CIRS clusters are comparable to the existing estimates.
In particular, our results exhibit good agreement with the bounds found by the analysis of the two massive clusters in \cite{2025A&A...698A..83P}, overlapping with both MACS 1206 and Abell S1063 within the quoted uncertainties. Moreover, we also recover a good agreement with the best-fit value of the vacuum permittivity for elliptical galaxies, whereas our results exhibits a tension ($\gtrsim 2\sigma$) with the estimate from disk galaxies, with a significance depending on the bin and gas prescription.

The comparison with the previous cluster-scale analysis should nevertheless be interpreted with caution. The constraints of \citet{2025A&A...698A..83P} were derived from two individual, massive, and relatively high-redshift clusters, using resolved gas and stellar mass profiles together with the stellar velocity-dispersion profile of the BCG. Our analysis instead considers lower-redshift stacked systems and relies on effective baryonic profiles reconstructed from scaling relations and from the average properties of the clusters in each bin. The comparison therefore combines possible intrinsic variations of the RG parameters with differences in cluster mass, radial coverage, and baryonic modelling.

Our analysis also shows a marked difference in the constraining power on the three RG parameters, compared to \cite{2025A&A...698A..83P}. The posterior distribution of $\epsilon_0$ is localized in all the runs, with best fit values of order $\epsilon_0\sim0.05$  for $\beta=2/3$
$\epsilon_0\sim0.2-0.3$, for $\beta=1$ whereas $\rho_c$ is only weakly bounded from below and $Q$ remains essentially unconstrained. This behaviour can be understood if most of the radial range considered in the kinematical analysis lies in the low-density regime $\rho_{\rm bar}\ll\rho_c$, where the gravitational permittivity approaches $\epsilon(\rho_{\rm bar})\simeq\epsilon_0$. In this regime, the dynamics constrain the asymptotic enhancement of the gravitational field but contain little information on the density at which the transition occurs or on its steepness. The exclusion of the innermost $50\,\mathrm{kpc}$ and the absence of an independently modelled BCG density profile may further reduce the sensitivity to the transition region. Consequently, the present analysis provides a test of the consistency of $\epsilon_0$ across different stacks, but cannot provide a meaningful test of the universality of $\rho_c$ and $Q$.

The constraints on $\epsilon_0$ obtained from the different
stacks are mutually compatible within the quoted uncertainties
and overlap with the previous cluster-scale estimates of
\citet{2025A&A...698A..83P}. In the present analysis, the gas
and stellar mass-profile parameters are varied jointly with
the RG parameters, so that the inferred constraints include
the uncertainty allowed by the adopted baryonic priors.
This consistency concerns $\epsilon_0$: the strongest tensions
reported in the previous study involved $\rho_c$ and $Q$,
which are not sufficiently constrained here to reassess
those discrepancies.

The two stacks selected with $A^2<0.7$ yield consistent
constraints on $\epsilon_0$ for both gas models. Moreover,
changing the gas-density slope produces smaller shifts in
their central estimates than in the three main stacks.
The agreement between the two selected velocity-dispersion
bins is therefore maintained across the two ICM prescriptions
tested here. This suggests a weaker sensitivity of their central estimates to the adopted gas model. Note that the Anderson--Darling statistic probes a specific aspect of the dynamical state through departures from Gaussianity in the line-of-sight velocity distribution. Since all bins are
analysed within the same modelling and inference framework, the coherence among the estimates of $\epsilon_0$ for the $A^2$-selected stacks supports
the use of this dynamical selection to mitigate possible systematics. However, the comparison with the full sample also involves changes in binning and in the reconstructed baryonic profiles, which may affect the resulting constraints.

The dependence on the ICM parametrization follows from the role of baryons in sourcing the RG gravitational field.
At radii much larger than the gas scale radius, the beta-model
density behaves as $\rho_{\rm gas}\propto r^{-3\beta}$.
For $\beta=2/3$, the density decreases as $r^{-2}$ and the
cumulative gas mass grows approximately linearly with radius;
for $\beta=1$, the density decreases as $r^{-3}$ and the
cumulative mass grows only logarithmically. These different
radial behaviours affect both the enclosed baryonic mass
and the local density entering the gravitational permittivity.
The shallower profile can support a more slowly declining acceleration and a flatter line-of-sight velocity-dispersion profile in the outskirts. The final prediction also depends on the fitted gas normalization and scale radius, the permittivity, and the orbital anisotropy. Although both gas prescriptions yield compatible constraints at the current precision, the shifts in the central estimates show that baryonic modelling remains an important source of uncertainty.

Finally, the Bayesian Information Criterion favours the Newtonian model with an NFW dark-matter component over the baryon-only RG model, with $\Delta\mathrm{BIC}>6$ in all the considered cases. The maximum likelihoods of the two models are nevertheless comparable; thus the BIC preference indicates that the additional parameter introduced in the RG analysis is not compensated by a corresponding improvement in the maximum likelihood. Moreover, since $\rho_c$ and $Q$ are only weakly identified by the data, the precise magnitude of the BIC difference should be interpreted with some caution. Indeed, if we repeat the analysis fixing, for instance, $Q = 3$, we do not find any statistical preference for one framework over the other.

\section{Conclusion}
\label{sec:conclusion}
In this work, we tested the phenomenological RG framework of \cite{2016arXiv160304943M} using the p.p.s distribution of 74 galaxy clusters from the CIRS sample \citep{2006AJ....132.1275R}. The clusters were separated into three bins of increasing velocity dispersion, stacked to reduce the impact of cluster asphericity and to increase the statistical power of the kinematical analysis. We also considered two additional stacks containing only clusters with an Anderson--Darling coefficient $A^2<0.7$, which is adopted as a proxy for relaxation. The gravitational potential and galaxy velocity-anisotropy profiles were reconstructed with the \texttt{MG-MAMPOSSt} code \citep{2021MNRAS.506..595P} in both a Newtonian framework with an NFW dark matter halo and a baryon-only RG framework.

Since resolved baryonic mass profiles are not available for the sample, we parametrised the gas and stellar mass-profiles using a $\beta$-model and an NFW, respectively. Their free parameters were inferred jointly with the dynamical parameters, using physically informed baryonic priors.
Two fixed gas-density slopes, $\beta=2/3$ and $\beta=1$, were considered to assess the sensitivity to ICM modelling.

Among the RG parameters, $\epsilon_0$ is appreciably
constrained, whereas $\rho_c$ is only weakly bounded
from below and $Q$ remains unconstrained. Across the
three main stacks, the central estimates of $\epsilon_0$
range from $0.04$ to $0.08$ for $\beta=2/3$ and from
$0.19$ to $0.37$ for $\beta=1$. The constraints are
mutually compatible within the quoted 95\% intervals
and overlap with previous cluster-scale estimates
\citep{2025A&A...698A..83P}. Nevertheless, the shifts
between the two gas prescriptions show that the inferred
permittivity remains sensitive to baryonic modelling.
The agreement concerns $\epsilon_0$ and does not establish
the universality of the full set of RG parameters.

The two stacks selected with $A^2<0.7$ yield compatible constraints on $\epsilon_0$ for both gas models and
show smaller shifts in their central estimates when the gas-density slope is changed. This provides an internal consistency check between the two selected
velocity-dispersion bins; however it does not produce statistically significant changes in the relative quality of the fits. The compatibility among the different stacks is encouraging, but cannot by itself establish that systematic effects associated with stacking and departures from equilibrium have been totally removed.

The Newtonian and RG models attain comparable maximum likelihoods. With all RG parameters free, however, the BIC favours the Newtonian model, with
$\Delta\mathrm{BIC}>6$, because the additional parameter is not accompanied by a sufficient improvement in fit.
Indeed, in the test with $Q=3$, the BIC shows no substantial preference between the two frameworks. The model comparison is therefore sensitive to the adopted parametrization, and its interpretation requires caution given the weak constraints on the transition parameters.

The main limitations of the present analysis arise from the construction of the baryonic mass profiles and from treating each stack as a single spherical system in dynamical equilibrium. The gas distribution is inferred indirectly from an $L_X$--$M_{\rm gas}$ scaling relation and calibrated using resolved profiles for only a small subset of the sample, while the stellar component is tied to the gas through an assumed universal galaxy-to-gas mass ratio. These assumptions are shared by all the stacks and therefore cannot be averaged down by stacking. 

In addition, neither the BCG nor the diffuse ICL are modelled explicitly. Although the ICM dominates the total baryonic mass of massive clusters, the BCG and ICL can constitute a substantial and halo-mass-dependent fraction of the stellar component, particularly in lower-mass systems \citep{Gonzalez2013,Kravtsov2018}. Their omission is especially relevant in RG, because the predicted acceleration depends on both the enclosed baryonic mass $M_{\rm bar}(<r)$ and the local density $\rho_{\rm bar}(r)$ entering the gravitational permittivity.  
Further progress will require resolved cluster-by-cluster gas-density and stellar-mass profiles, crucial to reduce the uncertainties, and an explicit treatment of BCG and ICL components.

Moreover, analyses of simulations will be necessary to specifically quantify how much stacking reduces cluster-specific systematics and to determine whether residual tensions may originate from baryonic and dynamical modelling uncertainties or from the stacking procedure itself. Programmes such as X-COP \citep{Ghirardini2019} and CHEX-MATE \citep{Arnaud2021} illustrate the level of resolved and homogeneous ICM information required for this purpose.

The new generation of wide-field multi-object spectroscopic surveys will greatly expand the available cluster p.p.s. information, although different programmes will provide very different sampling densities. While cluster galaxies are not a dedicated DESI target class, DESI spectroscopy already covers members of tens of thousands of groups and clusters and enables measurements of stacked velocity dispersions and projection effects; fibre allocation and target selection, however, limit the sampling of individual systems \citep{Myles2025}. The 4MOST eROSITA Galaxy Clusters Survey is designed to obtain spectroscopic redshifts for the optical counterparts of approximately $40~000$ groups and clusters and to sample more than 15 members per system at $z<0.7$ \citep{2019Msngr.175...39F}. This broad but relatively sparse coverage will be particularly valuable for ensemble dynamical calibration. In contrast, the 4MOST CHANCES survey will target approximately $5\times10^5$ galaxies around 100 massive clusters and two superclusters, extending to $5~r_{200}$ \citep{Sifon2025}. 
These targeted programmes will supply the sampling required for resolved p.p.s. analyses of individual clusters and for constructing better-controlled stacks. They can therefore make cluster kinematics a substantially more powerful and complementary probe of gravity, but increased statistics alone will not control the dominant systematics. Robust applications will still require the spectroscopic selection function, membership, dynamical state, and stacking procedure to be independently validated. Moreover, although kinematics is opening a bright era for the physics of galaxy clusters and for cosmology in general \citep{BPD2026}, its interpretation in baryon-coupled theories such as RG necessarily depends on external multi-wavelength constraints on the baryonic distribution.

\begin{acknowledgements}
LP acknowledges support by the Italian Ministry for Research and University (MUR) under 457 Grant ``Progetto Dipartimenti di Eccellenza 2023-2027'' (BiCoQ). AD acknowledges partial support from the grant InDark of the Italian
National Institute of Nuclear Physics (INFN). This research has made use of NASA’s Astrophysics Data System Bibliographic Services.
\end{acknowledgements}

\bibliography{References}
\appendix
\section{Example of marginal distributions in RG} \label{app:A}
\begin{figure}[htbp]
    \centering
    \includegraphics[width=1.5\linewidth]{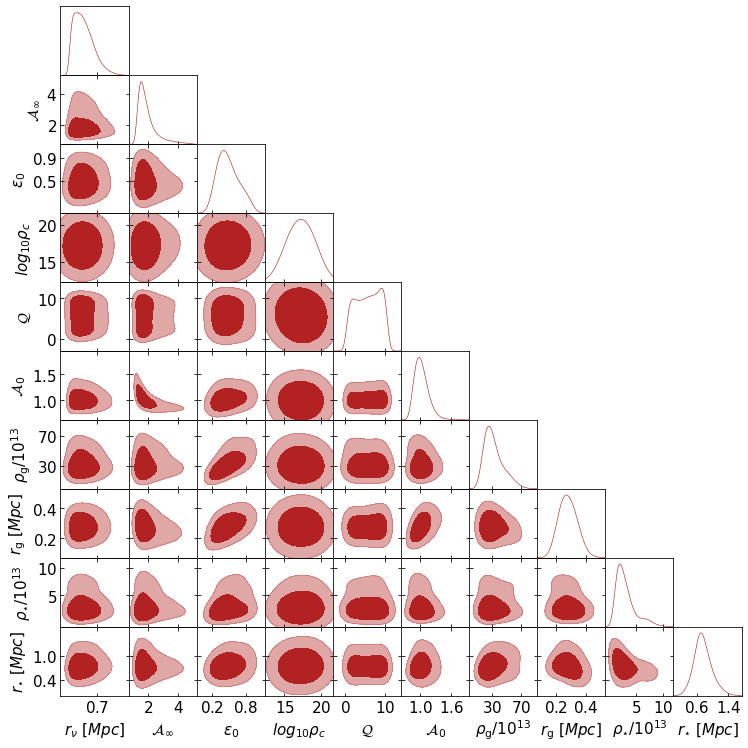}
    \caption{Marginal two-dimensional and one-dimensional distributions for the free parameters of the \textsc{MG-MAMPOSSt} run in RG for Bin 3, $\beta = 1$. The darker and lighter coloured areas represent the 68\% and 95\% confidence regions, respectively. The critical density and the baryonic densities are in units of $[{\text{M}_{\odot}/}{\text{Mpc}^3}]$ .}
    \label{fig:dist_bin3}
\end{figure}

In Fig. \ref{fig:dist_bin3} we show the marginalised distributions of the main parameters in our kinematical analysis in RG. As an illustrative example, the case of Bin 3 with $\beta = 1$, is plotted. No relevant correlations emerge among the RG and kinematical parameters, while  $\epsilon_0$ exhibits a slight positive correlation with the baryonic densities and radii. The same behaviour has been found in all the other cases.

\end{document}